\documentclass[preprint,amsmath,amssymb]{revtex4}

\usepackage{graphicx}

\newcommand{\beq}{\begin{equation}}
\newcommand{\eeq}{\end{equation}}
\newcommand{\beqa}{\begin{eqnarray}}
\newcommand{\eeqa}{\end{eqnarray}}

\newcommand{\ie}{{\it i.e.}}

\begin{document}

\title{FCNC Top Quark Decays in Extra Dimensions}
\author{G. A. Gonz\'alez-Sprinberg} 
\email{gabrielg@fisica.edu.uy}
\affiliation{Instituto de F\'{\i}sica, Facultad de Ciencias,
Universidad de la Rep\'ublica\\
Montevideo, Uruguay}
\author{R. Mart\1nez}
\email{remartinezm@unal.edu.co}
\author{J.-Alexis Rodriguez}
\email{jarodriguezl@unal.edu.co}
\affiliation{Departamento de F\'\i sica, Universidad Nacional de Colombia \\
Bogot\'a, Colombia}

%\date{}

\begin{abstract}
The flavor changing neutral top quark decay $t\rightarrow c
X$ is computed, where $X$ is a neutral standard model particle, in a extended
model with a single extra dimension. The cases for the
photon, $X = \gamma$, and a Standard Model Higgs boson, $X = H$, are
analyzed in detail in a non-linear $R_\xi$ gauge. We find that the
branching ratios can be enhanced by the dynamics originated in the
extra dimension. In the limit where $1/R >> m_t$, we have found $Br( t
\to c \gamma ) \simeq 10^{-10}$ for $ 1/R = 0.5 \, TeV$. For the
decay $t \to c H$, we have found $Br(t \to cH) \simeq 10^{-10}$ for
a low Higgs mass value. The branching ratios go to zero when $1/R
\to  \infty$.

\end{abstract}

\maketitle

\section{Introduction}

Flavor changing neutral currents (FCNC) are  very suppressed in
the standard model (SM): there are no tree level contributions and
at one loop level the charged currents operate with the Glashow-Iliopoulos-Maiani (GIM) mechanism.
The branching ratio for top quark FCNC decays into charm quarks are of
the order of $10^{-11}$ for $t\rightarrow c g$ and $10^{-13}$ for
$t\rightarrow c \gamma,(Z)$ in the framework of the
SM \cite{sm1,sm2}. This suppression can be traced back to the
loop amplitudes: they are controlled by down-type quarks, mainly
by the bottom quark, resulting in a $m_b^4/M_W^4$ factor which can
be compared to the enhancement factor that appears in the
$b\rightarrow s \gamma$ process where the top quark mass $m_t$ is involved
instead of $m_b$ in this factor. This
fourth power mass ratio is generated by the GIM mechanism and is
responsible for the suppression beyond naive expectations based on
dimensional analysis, power counting and Cabibbo-Kobayashi-Maskawa (CKM)-matrix elements involved. The
top quark decay into the SM Higgs boson is even more suppressed \cite{sm1,sm2}:
$Br(t \rightarrow c H ) \sim 10^{-13} - 10^{-15}$ for $M_Z \le M_H
\le 2 M_W$. These rates are far below the reach of any foreseen
high luminosity collider in the future. The highest FCNC top quark
rate in the SM is $t \rightarrow c g$, but this value is still six
orders of magnitude below the possibility of observation at the
LHC.
%All in all the detection of FCNC decays of the
%top quark at visible levels (viz. BR(t ¡ú cX) > 10.5) by the high
%luminosity colliders round the corner seems doomed to failure in
%the absence of new physics.

The discovery of these FCNC effects would be a hint of new physics
because of the large suppression in the SM.  These
FCNC decay modes can be strongly enhanced in scenarios beyond the
SM, where some of them could be even observed at the LHC or ILC. New
physics effects in  extended Higgs sector models,  SUSY and
left-right symmetric models were studied in references [1-5]. For example, in
various SUSY scenarios the  branching ratios  can go up to the
value $10^{-5}$ for the decay $t\rightarrow c g$. Also, virtual
effects of a $Z'$ gauge boson on these rare top quark decays were
studied. The decay $t \rightarrow c\gamma$ has been analyzed in reference
\cite{z1}, it has been shown that $B(t \rightarrow c \gamma)$ is at the $10^{-6}$
level in topcolor assisted technicolor models for $m_Z' = 1
\,TeV$, which would allow the detection of this process at future
 colliders.

%The $Z'$ contribution to $t->c\gamma$ was previously calculated in
%the context of topcolor-assisted technicolor (TC2) models [12].
%Those authors claim that, in that class of models, $B(t->c\gamma)$
%ranges between $1.3\times 10^{-6}$ and $1.7\times 10^{-7}$ for 1
% TeV      and 2 TeV.

 On the other hand, the use of
effective Lagrangians in parameterizing physics beyond the SM has
been studied extensively in FCNC top quark couplings and
decays \cite{el1,el2,el3}. This formalism generates a
model-independent parameterization of any new physics
characterized by higher dimension operators.
% In particular, the use of this method to
%place limits on new physics effects by studying their one-loop
%contributions to precisely measured observables has been proved to
%be effective in the study of anomalous couplings of vector gauge
%bosons [8, 9, 101], the top quark [6, 111] and the tao neutrino
%[121].
Under this approach, several FCNC transitions have been
also significantly constrained: $t\to c\gamma$ \cite{tosc, tcg}, $t\to cg$
\cite{tcg, cord}, $l_i\to l_j\gamma$ \cite{lar} and $H\to l_i l_j$ \cite{dia}.

New physics effects have also been introduced in models with large
extra dimensions (ED) \cite{ed}. In recent years, these models have been a
major source of inspiration for beyond the SM physics in the
ongoing research. In these scenarios the four dimensional SM
emerges as the low energy effective theory of models living in
more than four dimensions, where these extra dimensions are
orbifolded. The presence of infinite towers of Kaluza-Klein (KK)
modes are the remanent of the extended dimensional dynamics at low
energies. The size of the extra dimensions can be unexpectedly
large, with $1/R$ at the scale of a few $TeV$ without
contradicting the present experimental data \cite{eds}. Then, if these
KK-modes are light enough, they
 could be produced in the near future at the next
 generation of colliders. Scenarios where all the SM fields, 
fermions as well as bosons, propagate in the bulk are known as "universal extra dimensions"\cite{ued,pom}. In these theories the number
of KK-modes is conserved at each elementary vertex and
 the coupling of any excited KK-mode to two
zero modes is prohibited. Then the constraints on
the size of the extra dimensions obtained from the SM
precision measurements are less stringent than in the case
 where there is no conservation of the KK particles (non universal extra
 dimensions).

The impact of the new physics coming from UED models has been widely studied and constraints on the parameter $1/R$ have been obtained. Analysis of the precision electroweak observables led to the lower bound $1/R \gtrsim 700-800$ GeV for a light Higgs boson mass and to $1/R \gtrsim 300-400$ GeV for a heavy Higgs boson mass \cite{russell}. On the other hand, using the process $b \to s \gamma$, the resulting bound on the inverse compactification radius is $1/R \gtrsim 250$ GeV \cite{bsg}. Moreover, a recent analysis making use of the exclusive branching ratio $B \to K^* \gamma$ shows that under conservative assumptions $1/R \gtrsim 250$ GeV \cite{bkg}. And from the inclusive radiative $\bar B \to X_s \gamma$ decay, a lower bound on $1/R \gtrsim 600$ GeV at $95\%$ C.L. can be obtained and it is independent of the Higgs boson mass value \cite{bxs}. Contributions from UED models have been considered on several FCNC processes,  reference \cite{burasydemas} has found that the processes $K_L \to \pi^0 e^+ e^-, \Delta M_s, K^+ \to \pi^+ \nu \bar \nu, K_L \to \pi^0 \nu \bar \nu, B_{d,s} \to X_{d,s} \nu \bar \nu, K_L \to \mu^+ \mu^-, B_{d,s} \to \mu^+ \mu^-, B \to X_s \mu^+ \mu^-$ for $1/R\approx 300$ GeV are enhanced  relative to the SM expectation and the processes $B \to X_s \gamma, B\to X_s g, \epsilon'/\epsilon$ are suppressed respect to the SM. In general, the present data on FCNC processes are consistent with $1/R\gtrsim 300$ GeV \cite{burasydemas, report}. Exclusive $B \to K^* l^+ l^-,B \to K^* \nu \bar \nu$ and $B_s$ decays \cite{bkg, raros1} have been studied in the framework of the UED scenario and also rare semileptonic $\Lambda_b$ decays \cite{raros2}.

In this paper we study the FCNC decays of the top quark in a 
 universal extra dimension theory where all the SM fields live in five dimensions. In particular, we compute
the $t\rightarrow c \gamma$ and $t\rightarrow c H$ decay modes in
a non linear $R_\xi$. This gauge has the advantage of a reduced
number of Feynman diagrams as well as simplified  Ward identities.
These facts facilitates and clarify the calculation.

The paper is organized as follows: in section II we present the
general framework for the five dimensional Lagrangian and derive
the corresponding four dimensional Lagrangian and Feynman rules.
In section III and IV we compute the  decays mode $t\rightarrow c
\gamma$ and  $t\rightarrow c H$ respectively, and discuss the
hypothesis implicit in the calculation. Finally, in section V we
present some conclusions. In the Appendix (section VI) we show
 the terms in the Lagrangian that
 are important for the Feynman rules in our calculation.

\section{The Model}

We begin presenting  the SM Lagrangian in five dimensions; let $x = 0,1
,2, 3 $ be the normal coordinates and $x^4 = y$ the fifth one. The fifth extra dimension
is compactified on the orbifold 
 $S^1/Z_2$ orbifold of size $R$ which is the compatification radius.
%The usual ordinary four dimensional space
%is the one determined by the brane $y = 0$.
We consider a generalization of the SM where the fermions, the
gauge bosons and the Higgs doublet propagate in the five dimensions.
The Lagrangian $L $ can be written as
\begin{equation}  L = \int d^4x \,dy\, ({\cal L}_A + {\cal L}_H + {\cal L}_F
+ {\cal L}_Y) \label{lag}\end{equation} with
\begin{eqnarray} {\cal L}_A &=& - \frac{1}{4} W^{M N a}
W^a_{M N} - \frac{1}{4} B^{M N} B_{M N},
\nonumber\\
{\cal L}_H &=& (D_M \Phi)^\dagger D^M\Phi -V(\Phi),
\nonumber\\
{ \cal L}_F &=&  \,\left[\,\overline{Q} (i \Gamma^M D_M)Q+
\overline{U} (i \Gamma^M D_M) U +
\overline{D} (i \Gamma^M D_M) D\right],\nonumber\\
{\cal L}_Y &=& - \overline{Q} \widetilde{Y}_u \Phi^c U -
\,\overline{Q} \widetilde{Y}_d \Phi D + h.c. \label{lagg}
 \end{eqnarray}
The numbers $M, N = 0, 1, 2, 3, 5$ denote the five
dimensional Lorentz indexes, $W^a_{M N} = \partial_M
W^a_N -
\partial_N W^a_M + \widetilde{g} \epsilon^{abc} W^b_M W^c_N
$ is the strength field tensor for the $SU(2)_L$ electroweak gauge
group and $B_{M N} = \partial_M B_N -
\partial_N B_M$ is that of the $U(1)_Y$ group. The gauge fields
depend on $x$ and $y$. The covariant derivative is defined as
$D_M \equiv
\partial_M - i \widetilde{g}\, W^a_M T^a - i \widetilde{g}\, ' B_M Y$, where
$\widetilde{g}$ and $\widetilde{g}\,' $ are the five dimensional
gauge couplings constants for the groups $SU(2)_L$ and $U(1)_Y$
respectively, and $T^a$ and $Y$ are the corresponding generators.
The five dimensional gamma matrices $\Gamma_M$
 are $\Gamma_\mu = \gamma_\mu$ and $\Gamma_4 = i \gamma_5$ with the
 metric tensor given by $g_{M N} = ( +, -, -, -, - )$.
 The matter fields $Q$, $D$ and $U$ are fermionic four components
 spinors with the same quantum numbers as the corresponding SM
 fields. To simplify the notation we have suppressed the $SU(2)$
 and color indices. The standard and charge conjugate doublet
 standard Higgs fields are denoted by $\Phi(x,y)$ and
 $\Phi^c(x,y) = i \tau^2 \Phi^*(x,y)$;  $\widetilde{Y}_{u,d}$
are the Yukawa matrices in the five dimensional theory responsible
for the mixing  of different families whose indices were
suppressed in the notation for simplicity.
 We have not included in Eq.~(\ref{lagg}) the leptonic sector nor the
$SU(3)_c$ dynamics because it is not relevant  for our proposes.
%We assume that the fifth dimension is compactified in a circle of
%radius $R$ in which the points $y$ and $-y$ are identified as in a
%$S^1/Z_2$ orbifold.
The low energy theory will only have zero
modes for fields that are even under $Z_2$ symmetry: this is the
case for the Higgs doublet that we choose to be even under this
symmetry in order to have a standard zero mode Higgs field. The
Fourier expansions of the fields are given by:
\begin{eqnarray} 
B_\mu (x,y) &=& \frac{1}{\sqrt{\pi R}} B_\mu^{(0)}(x) +
\frac{\sqrt{2}}{\pi R}\sum_{n=1}^\infty B_\mu^{(n)}(x)
\cos\left(\frac{n y }{R}\right),
\nonumber\\
B_5 (x,y) &=& \frac{\sqrt{2}}{\pi R}\sum_{n=1}^\infty B_5^{(n)}(x)
\sin\left(\frac{n y }{R}\right),
\nonumber\\
 Q(x,y) &=& \frac{1}{\sqrt{\pi R}} Q_L^{(0)}(x) +
\frac{\sqrt{2}}{\pi R}\sum_{n=1}^\infty \left[Q_L^{(n)}(x)
\cos\left(\frac{n y }{R}\right) + Q_R^{(n)} \sin \left(\frac{n y
}{R}\right)\right],
\nonumber\\
U(x,y) &=& \frac{1}{\sqrt{\pi R}} U_R^{(0)}(x) +
\frac{\sqrt{2}}{\pi R}\sum_{n=1}^\infty \left[U_R^{(n)}(x) \cos\
\left( \frac{n y }{R}\right) + U_L^{(n)} \sin \left(\frac{n y
}{R}\right)\right],
\nonumber\\
D(x,y) &=& \frac{1}{\sqrt{\pi R}} D_R^{(0)}(x) +
\frac{\sqrt{2}}{\pi R}\sum_{n=1}^\infty \left[D_R^{(n)}(x) \cos\
\left( \frac{n y }{R}\right) + D_L^{(n)} \sin \left(\frac{n y
}{R}\right)\right].
\end{eqnarray}

The expansions for $B_\mu$ and $B_5$ are similar to the expansions
for the gauge fields and the Higgs doublet (but this last one
without the $\mu$ or $5$ Lorentz index). It is by integrating the fifth $y$
component in Eq.~(\ref{lag})
 that we obtain the usual interaction terms and the KK spectrum for ED models.

The interaction terms relevant for our calculation will be written
in a non-linear $R_\xi$ gauge (see for details \cite{des} and the
first reference in \cite{esm3}). For example, in this gauge there
is no mixing between  the gauge bosons and the charged and
neutral unphysical Higgs fields. Besides,   the interaction terms
are simplified in such a way that there are no trilinear terms
such as $W_\mu^+ G_W^- A_\mu$, where $G_W^-$ is an unphysical
Higgs field. We are interested in the third family of quarks
and $Q_t^{(n)}$ and $Q_b^{(n)}$ are the upper and lower
parts of the doublet $Q$. Similarly, the $U_t^{(n)}$ and
$D_b^{(n)}$ are the KK modes of the usual right-handed singlet
top and bottom quarks, respectively. There is a mixing between the mass and gauge
eigenstates of the KK top quarks ($Q_t^{(n)}$ and $U_t^{(n)}$)
where the mixing angle is given by $
 \tan (2 \alpha_t^n) =
m_t/m_n \label{mix}$ with $m_n = n/R$. For the $b$ quark the
mixing is quite similar, but at leading order the only masses that
remain are $m_t$ and $m_n$ and in this limit the mixing angle is
zero. This leads to the spectrum $m_{Q_b^n} = m_n$ and
$m_{Q_t^n} = m_{U^n} = \sqrt{m_t^2+m_n^2}$ for the excited modes
of the third family.  After dimensional reduction, the fifth
components of the charged gauge fields, $W_5^{-(n)}$, mix with the
KK modes of the charged component $\phi^{-(n)}$ of the Higgs
doublet. The unmixed states are thus the charged physical boson
$G_W^{-(n)}$ excited state and a would be Goldstone boson
$\phi_G^{-(n)}$ that contribute to the mass of the KK gauge
bosons:
\begin{eqnarray}
\phi_G^{\pm(n)} & = & \frac{m_n W_5^{\pm(n)}+i M_W
\phi^{\pm(n)}}{\sqrt{m_n^2 +
M_W^2}} \stackrel{\stackrel{n\not=0}{M_W\to 0}}{\longrightarrow} W_5^{\pm(n)}~, \nonumber\\
G_W^{\pm(n)} & = & \frac{i M_W W_5^{\pm(n)} + m_n
\phi^{\pm(n)}}{\sqrt{m_n^2 + M_W^2}}
\stackrel{\stackrel{n\not=0}{M_W\to 0}}{\longrightarrow}
\phi^{\pm(n)} \label{eq:physicalHiggs}~.
\end{eqnarray}

The final expression for the lagrangian can be found in the
Appendix A, where we show the terms that contribute to the decays we are
interested in.

\section{The $t\rightarrow c \, \gamma$ decay rate}

In this section, we present the calculation at one-loop level of the $t \to c \gamma$ process in the framework of a 5-dimensional universal ED model. We start with a naive calculation comparing the decay widths calculate in the SM and ED model and assuming that in the ED model only the third generation is running in the loop. 
The one loop SM width for the top quark decay into a charm quark plus a
gauge boson  can be approximated by
\beq 
\Gamma ( t \to c V)
\simeq |V_{bc}|^2 \,\alpha\,\alpha_{em}^2 \,m_t \,
\left(\frac{m_b}{M_W}\right)^4\,  \left(
1-\frac{m_V^2}{m_t^2}\right)^2 
\eeq 
where for a photon, the
neutral gauge boson  and a gluon we have  $\alpha = \alpha_{em}$
($V=\gamma, Z$) or $\alpha = \alpha_s$ ($V = g$) respectively.
These results can be compared to the ones expected for extra
dimensions, where the ratio $m_b/M_W$ is replaced by
$M_W/m_{n}$. Using these approximations we can naively estimate
the ratio,
\beq
\frac{\Gamma ( t \to c \gamma)_{ED}}{\Gamma (t \to
c \gamma)_{SM} } \simeq \left[ \frac{\left(\sum_n (M_W/m_n)^2\right)^2}{\left(m_b/M_W\right)^4}\right] . \label{ratio}
\eeq 
The sum on the KK tower of excited states can be evaluated as we will explain
later in the text and we obtain 
\beq \frac{\Gamma ( t \to c
\gamma)_{ED}}{\Gamma (t \to c \gamma)_{SM} }
 = \frac{\pi^4}{36}\,\left[
\frac{\left(M_W/(1/R)\right)}{ (m_b/M_W)}\right]^4 \simeq
 1.2 \times 10^{2}
\label{numero}
\eeq 
for $R^{-1} \sim 0.5\,  TeV$. We have already mentioned that the SM
 prediction for the branching fraction for the decay $t \to c \gamma$ is
 of the order
 $Br (t \to c \gamma) \sim 10^{-12} $. Then, from Eq.~(\ref{numero}) the
 branching fraction for ED models is $Br (t \to c\gamma )_{ED} \sim 1 \times 10^{-10}$ for
$R^{-1} = 0.5$ TeV.

The naive result on the $\Gamma(t \to c \gamma)_{ED}$ motivates a complete analysis of 
the one loop amplitude in extra dimensions. The general transition $q_i\rightarrow q_j + \gamma$
for arbitrary quark flavors $i, j$ in a non linear $R_\xi$ gauge
was studied in reference \cite{des}, where it was found that a reduced
number of Feynman diagrams as well as simplified Ward identities
greatly facilitates the calculation in this $R_\xi$ gauge.

For on-shell quarks and real photons the transition matrix element
is given by
\beq
M_\mu = i\sigma_{\mu\nu} k^\nu (F_2^L m_c P_L + F_2^R m_t
P_R),
\eeq
where $k_\mu$ is the photon momentum, $P_{R,L} =
(1\pm\gamma_5)/2$ and the magnetic transition form factors
$F_2^{L,R}$ are 
\beq
2 F_2^L \,m_c = ( B_1 + B_3 + 2 B_5
) m_t - A_3 m_c + A_{12} - 2 A_{11}, \eeq 
\beq 2 F_2^R \,m_t = (
A_1 + A_3 + 2 A_5 ) m_t - B_3 m_c + B_{12} - 2 B_{11} ,
\eeq
where the $A_i, B_i$ form factors are gotten from the most general
 Lorentz structure of the renormalized proper vertex \cite{des}.

Electromagnetic gauge invariance restricts the amplitude of this
decay to the form
\begin{equation}
{\cal M}(t \to c\gamma) =  \frac{i e g^2}{2^6 c^2_W \pi^2 m_t^2}\,
\epsilon_\alpha^* \,k_\mu \,(2 \widetilde{F}_{2}^R m_t)\,
\bar{u}_c \sigma^{\alpha\mu} P_R u_t
\label{amplitude}
\end{equation} 
where $c_W$ is the cosine of the weak mixing angle and the form
 factors $F_2^R$ and $\widetilde{F}_2^R$ are related by
 \beq
 F_2^R m_t = \frac{e g^2}{32 c_W^2 \pi^2\, m_t} \widetilde{F}_2^R .
 \label{f2wf2}
 \eeq
 The decay width for this process can be written as
 \beq 
\Gamma ( t\rightarrow c\,\gamma) =
 \frac{\alpha^3 m_t}{2^9 \pi^2 s_w^4 c_w^4}
|\widetilde{F}_2^R|^2 \approx 4.8 \times 10^{-7}
|\widetilde{F}_2^R|^2  . \label{width}
\eeq

In order to perform the one loop calculation, we consider two
scenarios. The first one, when the mass of the excited states associated to
the quarks from the three low-energy families are quasi-degenerated
at tree level, without any kind of radiative corrections to KK masses. In this case, when the excitations coming from the other quarks are taken into account, the transition amplitude for the process
$t \to c \gamma$ takes the form
\begin{eqnarray}
& \propto & \sum_{i=d,s,b} V_{ti}V_{ci}^* \frac{1}{(n/R)^2+m_{i}^2} \nonumber \\
&\approx& V_{tb}V_{cb}^* \frac{1}{(n/R)^2}\frac{m_{b}^2}{(n/R)^2}
\end{eqnarray}
where the last line can be obtained using the unitarity of the CKM
matrix and considering that the electroweak corrections of the first
two families are of order zero. $m_i$ is the mass of the down quark running into the loop. 
Therefore in this scenario, we notice that the naive expectation of the decay width $\Gamma(t\to c\gamma)$ given by (\ref{numero}) is suppressed by the factor
$(m_{b}/(n/R))^4$, and then the final result, including the KK states, is smaller than the SM value.

In the second scenario, we consider
that the most important contribution to the loop correction comes
from the excited KK states associated to the third generation. This
is a more realistic scenario because there is a mass hierarchy 
in the KK states from the different families, such as at low
energy. We should mention that in universal extra dimension theories, the fixed
points from the orbifold break the translational symmetry of the
extra dimension and it is possible to introduce new interactions
on the branes. In these new interactions, there are counterterms
that cancel the divergences of the radiative corrections, mass
terms, and mixing terms from the different family KK modes
\cite{fixpoint}. All our results are presented in the context of
this scenario.

Some of the Feynman rules for the model of Section II can be found
in the Appendix where all the relevant terms are shown. In Figure 1,
we illustrate the topology of the one-loop diagrams that are contributing.

\begin{figure}[h]
\includegraphics[scale=0.6]{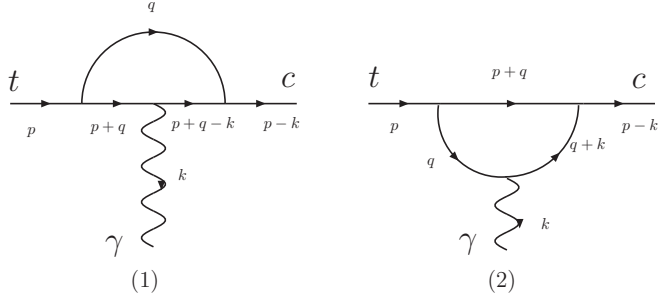}
\caption{\label{fig:1}Diagrams contributing to the $t\rightarrow
c\, \gamma$ decay in the non linear $R_\xi$ gauge.}
\end{figure}

In all the decays we are interested in, we neglect corrections of
order $(m_c/M_W)^2$. The mixing angle  between the gauge and mass
eigenstates for the KK excitations of the quarks are written in
section \ref{mix} and are zero for the leading order approximation. 
Other possible contributions are neglected
due to the Yukawa coupling constants. For example, diagrams
involving $G_W^{-(n)} \bar{Q}_{b}^{(n)} c_R^{(0)}$ are
proportional to the Yukawa coupling $\lambda_c$, and therefore to
$m_c^2$. Diagrams with $G_W^{-(n)}\, \bar{D}_b^{(n)}\,
t_L^{(0)}$ in the loop are proportional to $\lambda_b^2$ and then
to $m_b^2$ and can be neglected.

The leading contributions of type 1 diagrams (see
figure 1) to the decay come from the following particles circulating
in the loop:
\beq W_5^{(n)}Q_b^{(n)}Q_b^{(n)},\;\;\;
W_\mu^{(n)}Q_b^{(n)}Q_b^{(n)},\;\;\;
G_W^{(n)}Q_b^{(n)}Q_b^{(n)},\;\;\; G_W^{(n)}D_b^{(n)}D_b^{(n)}.
 \label{dia1}
\eeq
where the external photon is coupled to the fermion in the loop.
The sum of all these diagrams gives, for the form factor $F_2^R$,
the following expression
\beqa 
2 F_2^R &=&  \frac{2}{3} g^2e\,
V_{tb} V_{cb}^* \, \frac{i}{16\pi^2} \,
\int_0^1\,dz\int_0^{1-z}\,dw \frac{1-w-z}{\widetilde{X}} \nonumber\\
&&\times \left[z+(1-w)+z \frac{m_b^2}{M_W^2} \right] ,
\label{contr} 
\eeqa
where the factor $\widetilde{X}$ comes from the dimensional regularization
tricks of the product of inverse propagators of the particles
circulating in the loop
\beq 
\widetilde{X} =m^2_t z^2 + m_t^2 w z
- m_t^2 z +m_n^2 .
\eeq

The main contribution of type 2 diagrams (see figure 1) is the one with KK
excitation of the standard model gauge boson or a scalar field
circulating in the loop which are coupled to the external photon:
\beqa 
&&W_\mu^{(n)}W_\mu^{(n)}Q_b^{(n)},\;\;\;
W_5^{(n)}W_5^{(n)}Q_b^{(n)},\;\;\;
G_W^{(n)}G_W^{(n)}D_b^{(n)},\;\;\; G_W^{(n)}G_W^{(n)}Q_b^{(n)},
\nonumber\\&& (G_W^{(n)}W_\mu^{(n)}Q_b^{(n)}+
W_\mu^{(n)}G_W^{(n)}Q_b^{(n)}). 
\label{exc} 
\eeqa 
These terms contribute to the $F_2^R$ form factor with the following
expression 
\beqa 
2 F_2^R &=& - g^2 e \frac{i}{16\pi^2} \, V_{tb}
V_{cb}^* \, \int_0^1\,dz\int_0^{1-z}\,dw
\frac{1}{\widetilde{X}}\left\{ 4(2-3w-4z+2wz+2z^2)\right.
\nonumber \\
&-&(1-w-z)\left. \left[ (w+z)-(w+z-1/2) \frac{m_b^2}{M_W^2}
\right] \right\} .
\label{contr2} 
\eeqa
For a mass scale of the excited states much higher than the
electroweak scale, {\it i.e.}, $m_n \gg m_t$,  the
denominator can be approximated by $\widetilde{X} \simeq
m_n^2$. We also consider that the excited quarks rotate from
interaction to mass eigenstates with the same matrix of the
ordinary quarks. So, in the Yukawa lagrangian the interactions
$\bar{Q}_{bL}t_R^{(0)}G_W^{-(n)}$ and
$\bar{Q}_{bL}c_R^{(0)}G_W^{-(n)}$ are proportional to $V_{tb}$ and
$V_{cb}$ respectively, in the mass eigenstates.  If we compare the
leading contribution coming from these diagrams respect to
the SM contribution \cite{sm2}, it is 
\beq 
\frac{1}{m_n^2} \,:\,
\frac{m_b^2}{M_W^2}\frac{1}{M_W^2} .
\eeq
The numerical estimation of all these contributions is
straightforward. All the excited mass terms are proportional  to
$n/R$, except for the electroweak correction coming from the
symmetry breaking. From the numerical point of view this
correction does not change the results and can be neglected
without modifying the final estimates. Based on these hypothesis, we can
also take $\widetilde{X} \simeq m_n^2$ and, then,  the sum over
all the KK excited states can be easily done, as 
\beq \sum_n
\frac{1}{\widetilde{X}} \simeq \sum_n \frac{1}{\left(
\frac{n}{R}\right)^2} = \frac{\pi^2}{6}
\frac{1}{\left(\frac{1}{R}\right)^2} 
\eeq 
where in any numerical estimate  $1/R\simeq {\cal{O}}(1)$ TeV.

Thus, within this approximation, the sum over all the excited KK
states is equivalent to multiply the results obtained for the
first KK excited state by the factor $\pi^2/6 $. The sum of all contributions
using equations (\ref{contr}) and (\ref{contr2}) gives,
\begin{eqnarray}
2F_2^R&=&\frac{g^2e}{3}V_{tb}V_{cb}^*\frac{i}{16
\pi^2}\int_0^1\,dz\int_0^{1-z}\,dw
\frac{1}{\widetilde{X}}\left\{-22+35w-w^2 \right. \nonumber \\
&+&\left. 51z-30wz-29z^2+ (-4z^2-4wz+10z+6w+3)
\frac{m_b^2}{2M_W^2}\right\}
\nonumber \\
&\approx&\frac{g^2e}{18 m_n^2}V_{tb}V_{cb}^*\frac{i}{16
\pi^2}\left\{ -\frac{5}{2}+11 \frac{m_b^2}{M_W^2}\right\} .
\end{eqnarray}
By using equations (\ref{width}) and (\ref{f2wf2}), the numerical value for
the decay width is
\beq 
\Gamma (t \rightarrow c \gamma)
= 1.65 \times 10^{-10} \,GeV\;\; ,
\eeq 
for $R^{-1} = 0.5$ TeV, and  the branching fraction is 
\beq Br( t \rightarrow c \gamma )
\equiv \frac{\Gamma( t \rightarrow c \gamma)}{\Gamma( t
\rightarrow W b)} =1.08 \times 10^{-10}.
\eeq 
This result shows a branching ratio above the SM one, two orders of magnitude.

\section{The $t \rightarrow c \, H$ decay rate}

The invariant amplitude for the flavor changing decay of a top quark into a
charm quark plus a SM Higgs particle can be written
as
\beq 
{\cal M}( t \to c H ) = \overline{u}_c(p) \,\left( F_L
P_L + F_R P_R \right)\, u_t(k) ,
\eeq 
where $F_L$ and $F_R$ are form factors. In our notation we identify the external scalar
Higgs $H$ with the zero mode Higgs field $h^{(0)}$. From this amplitude we can compute the decay width
\beq
\Gamma ( t \to c H ) \,=\, \frac{m_t}{32
\pi}\, \left( 1-\frac{m_H^2}{m_t^2}\right)^2\,\left( |F_L|^2 +
|F_R|^2 \right) . \label{amplit}
\eeq

\begin{figure}[ht]
\includegraphics[scale=0.6]{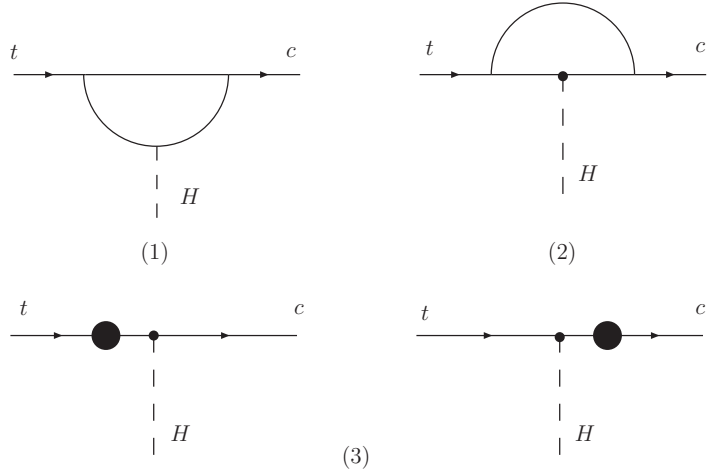}
\caption{\label{fig:2}Feynman diagrams for the $t\rightarrow
c\, H$ decay in extra dimensions.}
\end{figure}

The leading diagrams that contribute to the decay are shown in
Figure 2.
The leading group of type 1 diagrams in figure 2 is the one with KK
excitations of the SM quarks circulating in the loop
which are coupled to the external higgs:
\beq
W_\mu^{(n)}Q_b^{(n)}D_b^{(n)},\;\;\;
G_W^{(n)}Q_b^{(n)}Q_b^{(n)},\;\;\;
W_5^{(n)}Q_b^{(n)}D_b^{(n)} . \label{exch} 
\eeq
In this case, the
external Higgs is coupled to the excited quark $Q_b$ generating a
flavor changing $\bar{Q}^{(n)}D_bh^{(0)}$ (see the appendix), which
is proportional to the bottom quark mass $m_b$. The contributions
to the $F_L$, $F_R$ form factors are of the order of zero at
leading order,  $\ie \, F_L=F_R\approx 0$.

The leading diagrams of type 2 in figure 2 is the one with KK
excitations of the standard model gauge bosons and scalar fields
circulating in the loop which are coupled to the external higgs
boson $h^{(0)}$:
\beqa 
&&W_5^{(n)}W_5^{(n)}Q_b^{(n)},\;\;\;
G_W^{(n)}G_W^{(n)}Q_b^{(n)},\;\;\;
W_\mu^{(n)}W_\mu^{(n)}Q_b^{(n)}, \;\;\;
W_\mu^{(n)}G_W^{(n)}Q_b^{(n)}, \nonumber\\&&
G_W^{(n)}W_\mu^{(n)}Q_b^{(n)},\;\;\;
(W_5^{(n)}G_W^{(n)}Q_b^{(n)}+G_W^{(n)}W_5^{(n)}Q_b^{(n)})\, ,
\label{exchh} 
\eeqa 
and these contribute to the form factor with
the following expressions:
\beqa  
F_L &=&  \, g^3 V_{tb}
V_{cb}^*\frac{i}{16\pi^2} \, \int_0^1\,dz\int_0^{1-z}\,dw
\frac{1}{\widetilde{X}}\left\{ 2(1-w)m_t M_W \right.\nonumber \\
&-& \left.
\left[(z-1)(w+z)m_t^2+w(z-2)m_h^2\right]\frac{m_t}{m_W}\right\} 
\nonumber \\
 F_R &=&  g^3 V_{tb} V_{cb}^*\frac{i}{16\pi^2}
\, \int_0^1\,dz\int_0^{1-z}\,dw \frac{1}{\widetilde{X}}\left\{ w
m_t M_W  \right\} .
 \label{exc1} \eeqa
After evaluating the parametric integrals, the form factors are
given by 
\beqa 
F_L &=& \frac{g^3}{6 m_n^2}  \, V_{tb} V_{cb}^* \,
\frac{i}{16\pi^2}\left\{ 8 M_W m_t +\frac{5 m_t^3}{M_W}+ \frac{7
m_h^2 m_t}{2 m_W}\right\} ,
\nonumber \\
 F_R &=& \frac{g^3}{6 m_n^2} \,
V_{tb} V_{cb}^* \,\frac{i}{16\pi^2}  \left\{ M_W m_t \right\}.
\eeqa
Finally, the type 3 diagrams in figure 2 coming from a renormalized flavor
changing fermion line, do not contribute at leading order to the
decay width. The first one of these diagrams is proportional to the charm quark
mass because of the Higgs coupling, and therefore is negligible. For a
similar reason, the second diagram has a top quark mass factor, but the
self-energy part introduces the charm quark mass and this contribution
is suppressed  respect to the leading order.

From these form factors and equation (\ref{amplit}), we compute the $t\to cH$ decay width. Finally, the branching ratio is $Br (t \to c H) = 1.08 \times 10^{-10}$, for $m_H = 120$ \,GeV .

\section{Conclusions}

We have computed the decay widths $\Gamma(t \to c\, \gamma)$ and $\Gamma(t \to c\, H)$ in
a universal extra dimension model with a single extra dimension, where we have considered
that the most important contribution to the loop correction comes
from the excited KK states associated to the third generation. The
results show a branching ratio that is above the SM one.
The branching ratios for these two decay widths are of the order of
$10^{-10}$.

There is a strong dependence on top quark mass $m_t$ in the
amplitude of the $t \to c H$ process, it is coming from the type 2 diagrams in
figure 2 with an excited scalar in the loop, resulting in a
 $m_t^3/M_W^3$ factor. When we take the limit $m_n \gg m_t$, we find that the decay widths for the $t
\to c\gamma, (H)$ processes are decoupled respect to the new  scale $1/R$ and they go to zero.
Considering that the excited states of the quarks are
quasi-degenerated and the unitarity of the CKM matrix, the amplitudes for the flavor 
changing decay $t-c$ due to the contribution
of the excited states of the quarks, are suppressed by the factor
$(m_{b}/(n/R))^2$, and the predicted values are smaller than
the SM predictions.

\section{acknowledgements}

R. Martinez acknowledge the financial support from Fundación Banco de la Republica.

%UDELAR-Uruguay, CLAF and HELEN Project from  ALFA-EC funds.

\section{Appendix}
The Lagrangian can be separated in different terms as in the
following sum:
\begin{equation}
L=\sum_{m=1,2,3}\sum_{n=1}^\infty L_m^{(n)} \label{a1} .
\end{equation}
After symmetry breaking the interaction terms are included in the
terms  $L_1^{(n)}$ up to $L_3^{(n)}$. The first one, $L_1^{(n)}$ is

\beq L_1^{(n)}=\frac{g
m_t}{\sqrt2M_W}\bar{Q}_{bL}^{(n)}t_R^{(0)}G_W^{-(n)} - \frac{g
m_b}{\sqrt2 M_W}\bar{D}_{bR}^{(n)}t_L^{(0)}G_W^{-(n)} + h.c.
\label{l1} \eeq
The second term, $L_2^{(n)}$ has the excited gauge boson-fermion interactions:

\begin{eqnarray}
L_2^{(n)}&=&\frac{2}{3}\, e
\,A_{\mu}^{(0)}\left[\bar{Q}_{t}^{(n)}\gamma^\mu
{Q}_{t}^{(n)}+\bar{u}_{t}^{(n)}\gamma_\mu\bar{u}_{t}^{(n)}\right]+\left\{
\frac{g}{\sqrt2}
W_{\mu}^{+(n)}\bar{Q}_{bL}^{(n)}\gamma^\mu t_{L}^{(0)} \right.\nonumber \\
&+&\left. i\,e \,A_\mu^{(0)}\left[G_W^{-(n)}\partial^\mu
G_W^{+(n)} + W_5^{-(n)}\partial^\mu W_5^{+(n)} \right]
+\frac{g}{\sqrt2}\bar{Q}_{bR}^{(n)} t_L^{(0)}W_5^{-(n)}
\right. \nonumber\\
&+& \left. 2\, e \,\frac{n}{R} A_\mu^{(0)}\left[ W^{-(n)}_\mu
W_5^{+(n)}- W^{+(n)}_\mu W_5^{-(n)}\right]+h.c.\right\} \label{l2}
\end{eqnarray}
%\newpage
And the third term has the neutral Higgs boson interactions:
\begin{eqnarray}
L_3^{(n)}&=&-\frac{g m_t}{2 M_W} \bar{Q}_{t}^{(n)} U_{t}^{(n)}
h^{(0)} - \frac{g m_b}{2 M_W} \bar{Q}_{b}^{(n)} D_{b}^{(n)}h^{(0)}
\nonumber \\
&+&g M_W W_5^{+(n)}W_5^{-(n)}h^{(0)} + g M_W
W_\mu^{+(n)}W_\mu^{-(n)}h^{(0)}
\nonumber \\
&-&\frac{g m_h^2}{2M_W} G_W^{+(n)}G_W^{-(n)}h^{(0)}\nonumber \\
&+& \left\{i\frac{g}{2}\frac{n}{R}W_5^{(n)}G_W^{+(n)}h^{(0)} +
i\frac{g}{2}W_\mu^{+(n)}h^{(0)}\partial^\mu G_W^{-(n)}
 + h.c. \right\}
\end{eqnarray}

\end{document}